\documentclass[12pt]{article}

\newlength{\dinwidth}
\newlength{\dinmargin}
\usepackage{amssymb}

\def\1#1{{\bf #1}}
\def\2#1{{\cal #1}}
\def\3#1{{\sl #1}}
\def\4#1{{\tt #1}}
\def\5#1{{\sf #1}}
\def\6#1{{\mathfrak #1}}
\def\7#1{{\mathbb #1}}

\newcommand{\be}{\begin{equation}}
\newcommand{\ee}{\end{equation}}
\newcommand{\ba}{\begin{array}}
\newcommand{\ea}{\end{array}}
\newcommand{\bea}{\begin{eqnarray}}
\newcommand{\eea}{\end{eqnarray}}
\newcommand{\bean}{\begin{eqnarray*}}
\newcommand{\eean}{\end{eqnarray*}}
\newcommand{\nn}{\nonumber}

\newcommand{\impl}{\Rightarrow}
\newcommand{\restr}{\upharpoonright}
\newcommand{\ol}{\overline}

\newcommand{\del}{\partial}

\newcommand{\qed}{\ \ $\blacksquare$}
\newcommand{\qft}{quantum field theory}
\newcommand{\qfts}{quantum field theories}

\newcommand{\npb}{Nucl. Phys. \1B}
\newcommand{\cmp}{Commun. Math. Phys. }

\newcommand{\rmp}{Rev. Math. Phys. }
\newcommand{\jfa}{J. Funct. Anal. }

\newtheorem{defin}{Definition}[section]
\newtheorem{lemma}[defin]{Lemma}
\newtheorem{prop}[defin]{Proposition}
\newtheorem{theorem}[defin]{Theorem}
\newtheorem{coro}[defin]{Corollary}
\newtheorem{conj}[defin]{Conjecture}

\newcommand{\bdefin}{\begin{defin}}
\newcommand{\blemma}{\begin{lemma}}
\newcommand{\bprop}{\begin{prop}}
\newcommand{\btheor}{\begin{theorem}}
\newcommand{\bcoro}{\begin{coro}}
\newcommand{\bconj}{\begin{conj}}
\newcommand{\edefin}{\end{defin}}
\newcommand{\elemma}{\end{lemma}}
\newcommand{\eprop}{\end{prop}}
\newcommand{\etheor}{\end{theorem}}
\newcommand{\ecoro}{\end{coro}}
\newcommand{\econj}{\end{conj}}

\newcommand{\poinc}{Poincar\'{e}}
\newcommand{\prf}{{\it Proof. }}
\newcommand{\rem}{{\it Remark. }}
\newcommand{\rems}{{\it Remarks. }}

\newcommand{\sectreset}[1]{\section{#1}\setcounter{equation}{0}}

\begin{document}\noindent
{\sf DESY 97-081} \hfill {\sf ISSN 0418-9833} \\
{\sf May 1997, revised September 1997} \hfill {\sf hep-th/9705019}

\begin{center} \vskip 14mm
{\LARGE Superselection Structure of Massive Quantum Field Theories}\\[3.2mm] 
{\LARGE  in $1+1$ Dimensions}\\
\vskip 15mm
{\large Michael M\"uger\footnote{Supported by the Studienstiftung des deutschen Volkes and the CEE.}\footnote{Present address: Dipartimento di Matematica, Universit\`{a} di Roma ``Tor Vergata'', Via della Ricerca Scientifica, 00133 Roma, Italy. Email: mueger@axp.mat.uniroma2.it}} \\[5mm]
{II.\ Institut f\"ur Theoretische Physik, Universit\"at Hamburg\\[1mm]
Luruper Chaussee 149, D--22761~Hamburg, Germany\\[4mm]
May 5, 1997}
\end{center} 
\vskip 10mm

\abstract{We show that a large class of massive quantum field theories in $1+1$ 
dimensions, characterized by Haag duality and the split property for wedges, does not 
admit locally generated superselection sectors in the sense of Doplicher, Haag and 
Roberts. Thereby the extension of DHR theory to $1+1$ dimensions due to Fredenhagen, 
Rehren and Schroer is vacuous for such theories. Even charged representations which are
localizable only in wedge regions are ruled out. Furthermore, Haag duality
holds in all locally normal representations. These results are applied to the theory
of soliton sectors. Furthermore, the extension of localized representations of a 
non-Haag dual net to the dual net is reconsidered.
It must be emphasized that these statements do not apply to massless
theories since they do not satisfy the above split property. In particular, it is known
that positive energy representations of conformally invariant theories are DHR 
representations.}

\sectreset{Introduction}
It is well known that the superselection structure, i.e.\ the structure of physically 
relevant representations or `charges', of \qfts\ in low dimensional spacetimes 
gives rise to particle statistics governed by the braid group and is described by 
`quantum symmetries' which are still insufficiently understood. The meaning of
`low dimensional' in this context depends on the localization properties of 
the charges under consideration. In the framework of algebraic \qft\ \cite{haag,K} 
several selection criteria for physical representations of the observable algebra have 
been investigated. During their study of physical observables obtained from a field 
theory by retaining only the operators invariant under the action of a gauge group (of 
the first kind), Doplicher, Haag and Roberts were led to singling out the class of
locally generated superselection sectors. A representation is of this type if it 
becomes unitarily equivalent to the vacuum representation when restricted 
to the observables localized in the spacelike complement of an arbitrary double cone
(intersection of future and past directed light cones):
\be \pi\restr\2A(\2O')\cong\pi_0\restr\2A(\2O')\quad \forall\2O\in\2K \label{dhr} .\ee
Denoting the set of all double cones by $\2K$ we consider a \qft\ to be defined by its 
net of observables $\2K\ni\2O\mapsto\2A(\2O)$. This is a map which assigns to each 
double cone a $C^*$-algebra $\2A(\2O)$ satisfying isotony:
\be \2O_1\subset\2O_2 \ \impl \ \2A(\2O_1)\subset\2A(\2O_2) .\ee
This net property allows the quasilocal algebra to be defined by
\be \2A = \ol{\bigcup_{\2O\in\2K}\2A(\2O)}^{\|\cdot\|} .\ee
The net is local in the sense that
\be [\2A(\2O_1),\2A(\2O_2)]=\{0\} \ee
if $\2O_1,\2O_2$ are spacelike to each other. The algebra $\2A(G)$ associated with an
arbitrary subset of Minkowski space is understood to be the subalgebra of $\2A$ generated
(as a $C^*$-algebra) by all $\2A(\2O)$ where $G\supset\2O\in\2K$.
Furthermore, the \poinc\ group
acts on $\2A$ by automorphisms $\alpha_{\Lambda,x}$ such that
\be \alpha_{\Lambda,x}(\2A(\2O))=\2A(\Lambda\2O+x) \quad \forall\2O .\ee
This abstract approach is particularly useful if there is more than one vacuum.

One requires of a physically reasonable representation that at least the translations
(Lorentz invariance might be broken) are unitarily implemented:
\be \pi\circ\alpha_x(A)=U_\pi(x)\pi(A)U_\pi(x)^* ,\ee
the generators of the representation $x\mapsto U(x)$, i.e.\ the energy-momentum 
operators, satisfying the spectrum condition (positivity of the energy). 
Vacuum representations are characterized
by the existence of a unique (up to a phase) \poinc\ invariant vector. Furthermore we
assume them to be irreducible and to satisfy the Reeh-Schlieder property, the latter
following from the other assumptions if weak additivity is assumed.
In the analysis of superselection sectors satisfying (\ref{dhr}) relative to a fixed 
vacuum representation one usually assumes the latter to satisfy Haag duality\footnote{
$\2M'=\{X\in\2B(\2H)|XY=YX\,\forall Y\in\2M\}$ denotes the algebra of all bounded 
operators commuting with all operators in $\2M$. If $\2M$ is a unital $*$-algebra then
$\2M''$ is known to be the weak closure of a $\2M$.}
\be \pi_0(\2A(\2O))' = \pi_0(\2A(\2O'))'' \quad \forall\2O\in\2K ,\label{duality}\ee
which may be interpreted as a condition of maximality for the local algebras.
In \cite{dhr3,dhr4}, based on (\ref{duality}), a thorough analysis of the structure of 
representations satisfying (\ref{dhr}) was given, showing that the category of these
representations together with their intertwiners is monoidal (i.e.\ there is a product 
or, according to current fashion, fusion structure), rigid (i.e.\ there are conjugates) 
and permutation symmetric. In particular, the Bose-Fermi alternative, possibly with 
parastatistics, came out automatically although the analysis started from observable, 
i.e.\ strictly local, quantities. A lot more is known in this situation (cf. \cite{dr2})
but we will not need that.
A substantial part of this analysis, in particular concerning permutation statistics
and the Bose-Fermi alternative, is true only in at least $2+1$ spacetime dimensions.
The generalization to $1+1$ dimensions, where in general only braid group statistics 
obtains, was given in \cite{frs1} and applied to conformally invariant theories
in \cite{frs2}. Whereas for the latter theories all positive energy representations are 
of the DHR type \cite{bmt}, it has been clear from the beginning that the criterion
(\ref{dhr}) cannot hold for charged sectors in gauge theories due to Gauss' law.

Implementing a programme initiated by Borchers, Buchholz and Fredenhagen proved 
\cite{bf} for every
massive one-particle representation (where there is a mass gap in the spectrum 
followed by an isolated one-particle hyperboloid) the existence of a vacuum 
representation $\pi_0$ such that
\be \pi\restr\2A(\2C')\cong\pi_0\restr\2A(\2C')\quad \forall\2C \label{cones} .\ee
Here the $\2C$'s are spacelike cones which we do not need to define precisely.
In $\ge 3+1$ dimensional spacetime the subsequent analysis leads to essentially the same
structural results as the original DHR theory. Due to the weaker localization properties,
however, the transition to braid group statistics and the loss of group symmetry occur
already in $2+1$ dimensions, see \cite{fg}. In the $1+1$ dimensional situation with 
which we are concerned here, spacelike cones reduce to wedges (i.e.\ translates of
$W_R=\{x\in\7R^2\ | \ x^1\ge |x^0|\}$ and the spacelike complement $W_L=W_R'$).
Furthermore, the arguments in \cite{bf} allow us only to conclude the existence of two a
priori different vacuum representations $\pi_0^L, \pi_0^R$ such that the restriction of 
$\pi$ to left handed wedges (translates of $W_L$) is equivalent to $\pi_0^L$ and 
similarly for the right handed ones. As for such representations, of course long well 
known as soliton sectors, an operation of composition can only be defined if the `vacua 
fit together' \cite{fre2}, there is in general no such thing as permutation or braid 
group statistics. For lack of a better name soliton representations with 
coinciding left and right vacuum, i.e.\ representations which are localizable 
in wedges, will be called `wedge representations (or sectors)'.

There have long been indications that the DHR criterion might not be applicable to
massive $2d$-theories as it stands. The first of these was the fact, known for some 
time, that the fixpoint nets of Haag-dual field nets with respect to the action of a 
global gauge group do {\it not} satisfy duality even in simple sectors, whereas this is
true in $\ge 2+1$ dimensions. This phenomenon has been analyzed thoroughly in \cite{mue1}
under the additional assumption that the fields satisfy the split property for wedges. 
This property, which is expected to be satisfied in all massive \qfts, plays an 
important role also in the present work which we summarize briefly. 

In the next section we will prove some elementary consequences of Haag duality and the 
split property for wedges (SPW), in particular strong additivity and the time-slice 
property. The significance of our assumptions for superselection theory derives 
mainly from the fact that they preclude the existence of locally generated superselection
sectors. More precisely, if the vacuum representation satisfies Haag duality and the SPW
then every irreducible DHR representation is unitarily equivalent to the vacuum 
representation. This important and perhaps surprising result, to be proved in Sect.\ 3,
indicates that the innocent-looking assumptions of the DHR framework are quite
restrictive when they are combined with the split property for wedges. Although this 
may appear reasonable in view of the non-connectedness of $\2O'$, our 
result also applies to the wedge representations which are only localizable in wedges 
provided left and right handed wedges are admitted. In Sect.\ 4 we will prove the 
minimality of the relative commutant for an inclusion of double cone algebras which, via
a result of Driessler, implies Haag duality in all locally normal irreducible 
representations. In Sect.\ 5 the facts gathered in the preceding sections will be 
applied to the theory of quantum solitons thereby concluding our discussion of the 
representation theory of Haag-dual nets. Summing up the results obtained so far, the
representation theory of such nets is essentially trivial. On the other hand, dispensing
completely with a general theory of superselection 
sectors including composition of charges, braid statistics and quantum symmetry for 
massive theories is certainly not warranted in view of the host of more or less 
explicitly analyzed models exhibiting these phenomena. The only way to accommodate these
models seems to be to relax the duality requirement by postulating only wedge duality. 
In Sect.\ 6 Roberts' extension of localized representations to the dual net will be 
reconsidered and applied to the theories considered already in \cite{mue1}, namely 
fixpoint nets under an unbroken inner symmetry group. In this work we will not attempt 
to say anything concerning the quantum symmetry question. 

\sectreset{Strong Additivity and the Time-Slice Axiom}
Until further notice we fix a vacuum representation $\pi_0$ (which is always faithful) 
on a separable Hilbert space $\2H_0$ and omit the symbol $\pi_0(\cdot)$, 
identifying $\2A(\2O)\equiv\pi_0(\2A(\2O))$. Whereas 
we may assume the algebras $\2A(\2O),\ \2O\in\2K$ to be weakly closed, for more 
complicated regions $X$, in particular infinite ones like $\2O'$, we carefully
distinguish between the $C^*$-subalgebra 
\be \2A(X)\equiv\ol{\bigcup_{\2O\in\2K, \2O\subset X} \2A(\2O)}^{\|\cdot\|} \label{AX}\ee
of $\2A\equiv\pi_0(\2A)$ and its ultraweak closure $\2R(X)=\2A(X)''$.

\begin{defin} An inclusion $A\subset B$ of von Neumann algebras is standard \cite{dl}
if there is a vector $\Omega$ which is cyclic and separating for $A, B, A'\wedge B$.
\end{defin}
Due to the Reeh-Schlieder property, the inclusion $\2A(\2O_1)\subset\2A(\2O_2)$ 
($\2R(W_1)\subset\2R(W_2)$) is standard whenever $\2O_1\subset\subset\2O_2$ 
($W_1\subset\subset W_2$), i.e.\ the closure of $\2O_1$ is contained in the interior of 
$\2O_2$. ($W_1\subset\subset W_2$ is equivalent to the existence of a double cone 
$\2O$ such that $W_1\cup W_2'=\2O'$.)

\begin{defin} An inclusion $A\subset B$ of von Neumann algebras is split \cite{dal}, 
if there exists a type-I factor $N$ such that $A\subset N\subset B$. A net of 
algebras satisfies the split property (for double cones) \cite{buchh}
if the inclusion $\2A(\2O_1)\subset\2A(\2O_2)$  is split 
whenever $\2O_1\subset\subset\2O_2$ .\label{defsplit}\end{defin}
The importance of these definitions derives from the following result \cite{dal,dl}:
\blemma Let $A\subset B$ be a standard inclusion. Then the following are equivalent:
\begin{enumerate}
\item[(i)] The inclusion $A\subset B$ is split.
\item[(ii)] The is a unitary $Y$ such that $Y\,ab'\,Y^*=a\otimes b',\ a\in A, b'\in B'$.
\end{enumerate}
\label{equiv}\elemma
\rems 1. The implication (ii)$\impl$(i) is trivial, an interpolating type I factor being
given by $N=Y^*(\2B(\2H_0)\otimes\11)Y$.\\
2. The natural spatial isomorphism 
$\2A(\2O_1)\vee\2A(\2O_2)' \cong \2A(\2O_1)\otimes\2A(\2O_2)'$ implied by the split
property whenever $\2O_1\subset\subset\2O_2$ clearly restricts to 
\be  \2A(\2O_1)\vee\2R(\2O_2') \cong \2A(\2O_1)\otimes\2R(\2O_2'). \label{dual2}\ee
As an important consequence, every pair of normal
states $\phi_1\in\2A(\2O_1)_*, \phi_2\in\2R(\2O_2')_*$ extends to a normal state
$\phi\in(\2A(\2O_1)\vee\2R(\2O_2'))_*$. Physically this amounts to a form of statistical
independence between the regions $\2O_1$ and $\2O_2'$.\\
3. We emphasize that in the case where Haag duality fails 
($\2A(\2O)\subsetneq \2A(\2O')'$), requiring (\ref{dual2}) whenever 
$\2O_1\subset\subset\2O_2$ defines a weaker notion of split property since one can 
conclude only the existence of a type I factor $N$ such that
$\2A(\2O_1)\subset N\subset\2A(\2O_2')'=\2A^d(\2O_2)$.

In $1+1$ dimensions (and only there, cf.\ \cite[p.\ 292]{buchh}) the split property may
be strengthened by extending it to wedge regions. 
In this paper we will examine the implications of the split property for wedges (SPW).
The power of this
assumption in combination with Haag duality derives from the fact that one obtains strong
results on the relation between the algebras of double cones and of wedges. Some of these
have already been explored in \cite{mue1}, where, e.g., it has been shown that the 
local algebras associated with double cones are factors.
We recall some terminology introduced in \cite{mue1}:
the left and right spacelike complements of $\2O$ are denoted by $W^\2O_{LL}$ and
$W^\2O_{RR}$, respectively. Furthermore, defining $W^\2O_L={W^\2O_{RR}}'$ and 
$W^\2O_R={W^\2O_{LL}}'$ we have $\2O=W^\2O_L\cap W^\2O_R$. 

Before we turn to the main subject of this section, we remark on the relation 
between the two notions of Haag duality which are of relevance for this paper. 
In \cite{mue1}, as apparently in a large part of the literature, it was implicitly
assumed that Haag duality for double cones implies duality for wedges, i.e.\
\be \2R(W)'=\2R(W') \quad \forall W\in\2W, \label{wedge-d}\ee
where $\2W$ is the set of all wedge regions. Whereas there 
seems to be no general proof of this claim, for theories in $1+1$ dimensions satisfying 
the SPW we can give a straightforward argument, thereby also closing the gap in 
\cite{mue1}. In view of remark 3 after Lemma \ref{equiv} the following definition of 
the split property for wedges is slightly weaker than the obvious modification of 
Def.\ \ref{defsplit}, but seems more natural from a physical point of view (cf.\ 
remark 2):
\begin{defin} A net of algebras satisfies the split property for wedges
if the map $x\otimes y\mapsto xy,\ x\in\2R(W_1), y\in\2R(W_2)$ extends to an isomorphism
between $\2R(W_1)\otimes\2R(W_2)$ and $\2R(W_1)\vee\2R(W_2)$ whenever 
$W_1\subset\subset W_2'$. By standardness this isomorphism is automatically spatial in
the sense of Lemma \ref{equiv}, (ii). In the case where $W_1=W^\2O_{LL},\,W_2=W^\2O_{RR}$
the canonical implementer \cite{dl} will be denoted $Y^\2O$. \label{spw}\edefin

\bprop Let $\2A(\2O)$ be a net of local algebras in $1+1$ dimensions, satisfying Haag
duality (for double cones) and the SPW. Then $\2A$ satisfies wedge duality and the
inclusion $\2R(W_1)\subset\2R(W_2)$ is split whenever $W_1\subset\subset W_2$. 
\label{wdual}\eprop
\prf Appealing to the definition (\ref{AX}), duality for double cones is clearly 
equivalent to
\be \2A(\2O)'=\2R(W^\2O_{LL})\vee\2R(W^\2O_{RR}) \quad \forall\2O\in\2K .\ee
Given a right wedge $W$, let $\2O_i,\ i\in\7N$ be an increasing sequence of double cones
all of which have the same left corner as $W$ and satisfying $\cup_i \2O_i=W$. Then
we clearly have $\2R(W)=\bigvee_i \2A(\2O_i)$ and
\be \2R(W)'=\bigwedge_i \2A(\2O_i)'=\bigwedge_i 
   \left(\2R(W')\vee\2R(W^{\2O_i}_{RR})\right) .\label{AX1}\ee
Using the unitary equivalence 
$Y^{\2O_1}\,\2R(W')\vee\2R(W^{\2O_1}_{RR})\,Y^{\2O_1*}=\2R(W')\otimes\2R(W^{\2O_1}_{RR})$
, the right hand side of (\ref{AX1}) is equivalent to 
\be \bigwedge_i \left(\2R(W')\otimes\2R(W^{\2O_i}_{R})\right) 
   = \2R(W') \bigotimes \bigwedge_i \2R(W^{\2O_i}_{R}) = \2R(W')\otimes\7C\11 ,\ee
where we have used the consequence $\wedge_i \2R(W^{\2O_i}_{R})=\7C\11$ of 
irreducibility. This clearly proves $\2R(W)'=\2R(W')$. The final claim follows from
Lemma \ref{equiv}. \qed

Now we are prepared for the discussion of additivity properties, starting with the easy
\blemma 
\bea \2R(W^\2O_{LL})\vee\2A(\2O)&=&\2R(W^\2O_L), \label{sa0}\\
   \2R(W^\2O_{RR})\vee\2A(\2O)&=&\2R(W^\2O_R).  \label{sa1}\eea
\label{a1}\elemma
\rem Equivalently, the inclusions 
$\2R(W^\2O_{LL})\subset\2R(W^\2O_L), \2R(W^\2O_{RR})\subset\2R(W^\2O_R)$ are normal.
\prf Under the unitary equivalence 
$\2R(W^\2O_{LL})\vee\2R(W^\2O_{RR})\cong\2R(W^\2O_{LL})\otimes\2R(W^\2O_{RR})$ we
have $\2R(W^\2O_{LL})\cong\2R(W^\2O_{LL})\otimes\11$ and 
$\2A(\2O)=\2R(W^\2O_L)\cap\2R(W^\2O_R)\cong\2R(W^\2O_R)\otimes\2R(W^\2O_L)$. Thus 
$\2R(W^\2O_{LL})\vee\2A(\2O)\cong(\2R(W^\2O_{LL})\vee\2R(W^\2O_R))\otimes\2R(W^\2O_L)$.
Due to wedge duality and factoriality of the wedge algebras this equals
$\2B(\2H_0)\otimes\2R(W^\2O_L)\cong\2R(W^\2O_L)$. We emphasize that all above 
equivalences are established by the same unitary transformation.
The second equation is proved in the same way. \qed\\
\rem The proof of factoriality of wedge algebras in \cite{dri1} relies, besides the
usual net properties, on the spectrum condition and on the Reeh-Schlieder theorem. 
This is the only place where positivity of the energy and weak additivity enter into
our analysis.

Consider now the situation depicted in Fig.\ \ref{f1}.
\begin{figure}
\[\ba{c}\begin{picture}(300,100)(-150,-50)\thicklines
\put(20,0){\line(-1,1){40}}
\put(20,0){\line(-1,-1){40}}
\put(-20,0){\line(1,1){40}}
\put(-20,0){\line(1,-1){40}}
\put(20,0){\line(1,1){40}}
\put(20,0){\line(1,-1){40}}
\put(-20,0){\line(-1,1){40}}
\put(-20,0){\line(-1,-1){40}}
\put(55,0){\line(-1,1){40}}
\put(55,0){\line(-1,-1){40}}
\put(55,0){\line(1,1){40}}
\put(55,0){\line(1,-1){40}}
\put(35,-5){$\tilde{\2O}$}
\put(-5,-5){$\2O$}
\put(-70,-5){$W^\2O_{LL}$}
\put(-30,15){$W^\2O_L$}
\end{picture} \ea\]
\caption{Double cones sharing one point}
\label{f1}\end{figure}
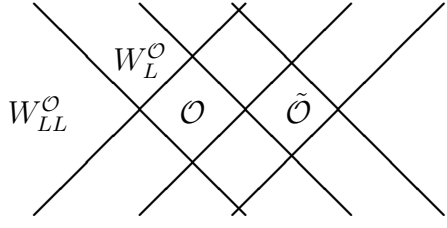
In particular, $\2O,\tilde{\2O}$ are spacelike separated double cones the closures of
which share one point. Such double cones will be called {\it adjacent}.
\blemma Let $\hat{\2O}=\sup(\2O,\tilde{\2O})$ be the smallest double cone
containing $\2O,\tilde{\2O}$. Then
\label{a2}\elemma
\be \2A(\2O)\vee\2A(\tilde{\2O})=\2A(\hat{\2O}) .\label{sa2}\ee
\prf In the situation of Fig.\ \ref{f1} we have 
$\hat{\2O}=W^\2O_R\cap W^{\tilde{\2O}}_L$.
Under the unitary equivalence considered above we have
$\2A(\tilde{\2O})\cong\11\otimes\2A(\tilde{\2O})$ since $\tilde{\2O}\subset W^\2O_{RR}$. 
Thus 
$\2A(\2O)\vee\2A(\tilde{\2O})\cong\2R(W^\2O_R)\otimes(\2R(W^\2O_L)\vee\2A(\tilde{\2O}))$.
But now $W^\2O_L=W^{\tilde{\2O}}_{LL}$ leads to 
$\2R(W^\2O_L)\vee\2A(\tilde{\2O})=\2R(W^{\tilde{\2O}}_L)$ via the preceding lemma.
Thus $\2A(\2O)\vee\2A(\tilde{\2O})\cong\2R(W^\2O_R)\otimes\2R(W^{\tilde{\2O}}_L)$
which in turn is unitarily equivalent to
$\2R(W^\2O_R)\wedge\2R(W^{\tilde{\2O}}_L)=\2A(\hat{\2O})$. \qed\\
\rem In analogy to chiral conformal field theory we denote this property {\it strong
additivity}.

With these lemmas it is clear that the \qfts\ under consideration are $n$-{\it regular}
in the sense of the following definition for all $n\ge 2$.
\bdefin A \qft\ is $n$-regular if 
\be \2R(W_1)\vee\2A(\2O_1)\vee\ldots\vee\2A(\2O_{n-2})\vee\2R(W_2)
  =\2B(\2H_0) ,\ee
whenever $\2O_i,\ i=1,\ldots,n-2$ are mutually spacelike double cones such that the sets
$\ol{\2O_i}\cap\ol{\2O_{i+1}},\ i=1,\ldots,n-3$ each contain one point and where
the wedges $W_1, W_2$ are such that
\be W_1\cup W_2=\left(\bigcup_{i=1}^{n-2} \2O_i \right)' .\ee
\edefin
\bcoro A \qft\ in $1+1$ dimensions satisfying Haag duality and the SPW fulfills the
(von Neumann version of the) time-slice axiom, i.e.\
\be \2R(S)=\2B(\2H_0) ,\ee
whenever $S=\{x\in \7R^2\,|\,x\cdot\eta\in(a,b)\}$ where $\eta\in\7R^2$ is timelike 
and $a<b$.
\ecoro
\prf The timeslice $S$ contains an infinite string $\2O_i,\,i\in\7Z$ of mutually
spacelike double cones as above. Thus the von Neumann algebra generated by all these 
double cones contains each $\2A(\2O),\,\2O\in\2K$ from which the claim follows by
irreducibility. \qed\\
\rems 1. We wish to emphasize that this statement on von Neumann algebras is weaker than
the $C^*$-version of the timeslice axiom, which postulates that the $C^*$-algebra
$\2A(S)$ generated by the algebras $\2A(\2O),\,\2O\subset S$ equals the quasilocal
algebra $\2A$. We follow the arguments in \cite[Sec. III.3]{haag} to the effect that 
this stronger assumption should be avoided.\\
2. It is interesting to confront the above result with the investigations concerning the
time-slice property \cite{gff} and the split property \cite[Thm. 10.2]{dl} in the context
of generalized free fields (in $3+1$ dimensions). In the cited works it was proved that
generalized free fields possess the time-slice property iff (roughly) the spectral
measure vanishes sufficiently fast at infinity. On the other hand, the split property 
imposes strong restrictions on the spectral measure, in particular it must be atomic 
without a accumulation point at a finite mass. The split property (for double cones) is,
however, neither necessary nor sufficient for the time-slice property.

\sectreset{Absence of Localized Charges}
Whereas the results obtained so far are intuitively plausible, we will now prove a
no-go theorem which shows that the combination of Haag duality and the SPW is extremely
strong.
\btheor Let $\2O\mapsto\2A(\2O)$ be a net of observables satisfying Haag duality and
the split property for wedges. Let $\pi$ be a representation of the quasilocal algebra
$\2A$ which satisfies
\be \pi\restr\2A(W)\cong\pi_0\restr\2A(W)\quad \forall W\in\2W ,\label{W}\ee
where $\2W$ is the set of all wedges (left and right handed). Then $\pi$ is equivalent 
to an at most countable direct sum of representations which are unitarily equivalent 
to $\pi_0$:
\be \pi=\bigoplus_{i\in I}\pi_i,\ \ \ \pi_i\cong\pi_0 .\ee
In particular, if $\pi$ is irreducible it is unitarily equivalent to $\pi_0$. 
\label{no-go}\etheor
\rem A fortiori, this applies to DHR representations (\ref{dhr}).\\
\prf Consider the geometry depicted in Fig.\ \ref{f2}.
\begin{figure}
\[\ba{c}\begin{picture}(300,150)(-150,-75)\thicklines
\put(40,0){\line(1,1){60}}
\put(40,0){\line(1,-1){60}}
\put(40,0){\line(-1,1){40}}
\put(40,0){\line(-1,-1){40}}
\put(0,0){\line(1,1){60}}
\put(0,0){\line(1,-1){60}}
\put(-40,0){\line(1,1){60}}
\put(-40,0){\line(1,-1){60}}

\put(60,-5){$W$}
\put(37,-25){$W_1$}
\put(15,-45){$W_2$}
\put(15,-5){$\2O_1$}
\put(-20,5){$\2O_2$}
\end{picture} \ea\]
\caption{A split inclusion of wedges}
\label{f2}\end{figure}
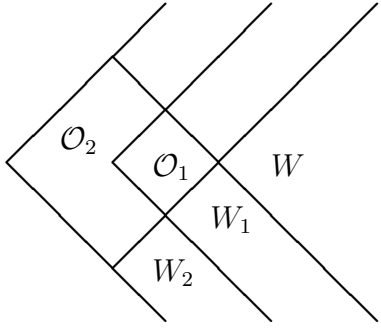
If $\pi$ is a representation satisfying (\ref{W}) then there is a unitary 
$V: \2H_\pi\to\2H_0$ such that, setting
$\rho=V\pi(\cdot)V^*$, we have $\rho(A)=A$ if $A\in\2A(W')$. Due to normality on
wedges and wedge duality, $\rho$ continues to normal endomorphisms of 
$\2R(W), \2R(W_1)$. By the split property there are type-I factors $M_1, M_2$ such that
\be \2R(W)\subset M_1\subset\2R(W_1)\subset M_2\subset\2R(W_2) .\ee
Let $x\in M_1\subset\2R(W_1)$. Then $\rho(x)\in\2R(W_1)\subset M_2$. Furthermore, 
$\rho$ acts trivially on $M_1'\cap\2R(W_2)\subset\2R(W)'\cap\2R(W_2)=\2A(\2O_2)$,
where we have used Haag duality. Thus $\rho$ maps $M_1$ into 
$M_2\cap(M_1'\cap\2R(W_2))'\subset M_2\cap(M_1'\cap M_2)'=M_1$, the last identity
following from $M_1, M_2$ being type-I factors. By \cite[Cor. 3.8]{lo1} every 
endomorphism
of a type I factor is inner, i.e.\ there is a (possibly infinite) family of isometries 
$V_i\in M_1,\ i\in I$ with $V_i^* V_j=\delta_{i,j},\sum_{i\in I}V_i V_i^*=\11$ such that 
\be \rho(A)=\eta(A)\quad\forall A\in M_1 ,\label{x}\ee
where
\be \eta(A)\equiv\sum_{i\in I}V_i\, A \, V_i^*, \ A\in\2B(\2H_0) .\label{y}\ee
(The sum over $I$ is understood in the strong sense.)
Now, $\rho$ and thus $\eta$ act trivially on 
$M_1\cap\2R(W)'\subset\2R(W_1)\cap\2R(W)'=\2A(\2O_1)$, which implies 
\be V_i\in M_1\cap(M_1\cap\2R(W)')'=\2R(W) .\ee
Thanks to Lemma \ref{a1} we know that for every wedge $\hat{W}\supset\supset W$
\be \2R(\hat{W})=\2R(W)\vee\2A(\2O) ,\ee
where $\2O=\hat{W}\cap W'$. From the fact that $\rho$ acts trivially on $\2A(W')$ it 
follows that (\ref{x}) is true also for $A\in\2A(\2O)$. By assumption, $\rho$ is normal
also on $\2A(\hat{W})$ which leads to (\ref{x}) on $\2A(\hat{W})$. As this holds for
every $\hat{W}\supset\supset W$, we conclude that
\be \pi(A)=\sum_{i\in I} V^*V_i\, A \, V_i^*V \quad \forall A\in\2A .\ee
\qed\\
\rems 1. The main idea of the proof is taken from \cite[Prop. 2.3]{dopl}. \\
2. The above result may seem inconvenient as it trivializes the DHR/FRS superselection 
theory \cite{dhr3,dhr4,frs1} for a large class of massive quantum field theories in 
$1+1$ dimensions. It is not so clear what this means with respect to field theoretical
models since little is known about Haag duality in nontrivial models. \\
3. Conformal \qfts\ possessing no representations besides the vacuum representation, 
or `holomorphic' theories, have been the starting point for an analysis of `orbifold'
theories in \cite{dvvv}. In \cite{mue1}, which was motivated by the desire to obtain a 
rigorous understanding of orbifold theories in the framework of massive two 
dimensional theories, the present author postulated the split property for wedges and
claimed it to be weaker than the requirement of absence of nontrivial representations.
Whereas this claim is disproved by Thm.\ \ref{no-go}, as far as localized (DHR or wedge)
representations of Haag dual theories are concerned, none of the results of \cite{mue1} 
is invalidated or rendered obsolete.

\sectreset{Haag Duality in Locally Normal Representations}
\begin{figure}
\[\ba{c}
\begin{picture}(300,150)(-150,-75)\thicklines
\put(0,60){\line(1,-1){100}}
\put(0,60){\line(-1,-1){100}}
\put(0,-60){\line(1,1){100}}
\put(0,-60){\line(-1,1){100}}

\put(0,20){\line(1,-1){60}}
\put(0,20){\line(-1,-1){60}}
\put(0,-20){\line(1,1){60}}
\put(0,-20){\line(-1,1){60}}

\put(-5,-5){$\2O$}
\put(-5,35){$\hat{\2O}$}
\put(-45,-5){$\2O_L$}
\put(35,-5){$\2O_R$}
\put(-95,-5){$W^{\hat{\2O}}_{LL}$}
\put(75,-5){$W^{\hat{\2O}}_{RR}$}
\put(-73,15){$W^\2O_{LL}$}
\put(47,15){$W^\2O_{RR}$}
\end{picture} 
\ea\]
\caption{Relative commutant of double cones}
\label{f3}\end{figure}
A further crucial consequence of the split property for wedges is observed in the 
following
\bprop Let $\2O\mapsto\2A(\2O)$ be a net satisfying Haag duality (for double cones)
and the split property for wedges. Then for every pair $\2O\subset\subset\hat{\2O}$
we have
\be \2A(\hat{\2O})\wedge\2A(\2O)' = \2A(\2O_L)\vee\2A(\2O_R) ,\ee
where $\2O_L, \2O_R$ are as in Fig.\ 3. \label{relcomm}\eprop
\prf By the split property for wedges there is a unitary
operator $Y^\2O:\2H_0\to\2H_0\otimes\2H_0$ such that 
$\2R(W^\2O_{LL})\vee\2R(W^\2O_{RR})=Y^{\2O*}(\2R(W^\2O_{LL})\otimes\2R(W^\2O_{RR}))Y^\2O$.
More specifically, 
\be Y^\2O\,xy\,Y^{\2O*}=x\otimes y \quad \forall x\in\2R(W^\2O_{LL}),\,y\in
  \2R(W^\2O_{RR}). \ee
By Haag duality 
$\2A(\2O)'=\2R(W^\2O_{LL})\vee\2R(W^\2O_{RR})\cong\2R(W^\2O_{LL})\otimes\2R(W^\2O_{RR})$
and $\2A(\hat{\2O})'=\2R(W^{\hat{\2O}}_{LL})\vee\2R(W^{\hat{\2O}}_{RR})$.
Now $\2R(W^{\hat{\2O}}_{LL/RR})\subset\2R(W^\2O_{LL/RR})$
implies $\2A(\hat{\2O})'\cong\2R(W^{\hat{\2O}}_{LL})\otimes\2R(W^{\hat{\2O}}_{RR})$ 
under the same equivalence $\cong$ provided by $Y^\2O$, and thus
\be \2A(\hat{\2O})\cong(\2R(W^{\hat{\2O}}_{LL})\otimes\2R(W^{\hat{\2O}}_{RR}))'=
   \2R(W^{\hat{\2O}}_R)\otimes\2R(W^{\hat{\2O}}_L) ,\ee
where we have used wedge duality and the commutation theorem for tensor products.
Now we can compute the relative commutant as follows:
\be\ba{ccccc} \2A(\hat{\2O})\wedge\2A(\2O)' & \cong & 
   (\2R(W^{\hat{\2O}}_R)\otimes\2R(W^{\hat{\2O}}_L)) & \wedge &
            (\2R(W^\2O_{LL})\otimes\2R(W^\2O_{RR})) \\
  &=& (\2R(W^{\hat{\2O}}_R)\wedge\2R(W^\2O_{LL})) & \otimes &
      (\2R(W^{\hat{\2O}}_L)\wedge\2R(W^\2O_{RR})) \\
  &=& \2A(\2O_L)\otimes\2A(\2O_R) & \cong & \2A(\2O_L)\vee\2A(\2O_R) .\ea\ee
We have used Haag duality in the form 
$\2R(W^{\hat{\2O}}_R)\wedge\2R(W^\2O_{LL})=\2A(\2O_L)$
and similarly for $\2A(\2O_R)$. \qed\\
\rems 1. Readers having qualms about the above computation of the intersection of tensor
products are referred to \cite[Cor. 5.10]{t}, which also provides the justification 
for the arguments in Sect.\ 2.\\
2. Recalling that $\2R(\2O)=\2A(\2O)$ and that the algebras of regions other than 
double cones are defined by additivity, (\ref{relcomm}) can be restated as follows:
\be \2R(\hat{\2O})\cap\2R(\2O)' = \2R(\hat{\2O}\cap\2O') .\label{relcomm1}\ee
In conjunction with the assumed properties of isotony, locality and Haag duality for
double cones (\ref{relcomm1}) entails that the map $\2O\mapsto\2R(\2O)$ is a homomorphism
of orthocomplemented lattices as proposed in \cite[Sect.\ III.4.2]{haag}. While the
discussion in \cite[Sect.\ III.4.2]{haag} can be criticized, the class of models 
considered in this paper provides examples where the above lattice homomorphism is in
fact realized.

The proposition should contribute to the understanding of Thm.\ \ref{no-go} as
far as DHR representations are concerned. In fact, it already implies the absence of
DHR sectors as can be shown by an application of the triviality criterion for local 
1-cohomologies \cite{r1} given in \cite{r4}, see also \cite{r-aqft}.\\
{\it Sketch of proof.} Let $z\in Z^1(\2A)$ be the local 1-cocycle associated according 
to \cite{r1,r4} with a representation $\pi$ satisfying the DHR criterion.
Due to Prop.\ \ref{relcomm} it satisfies 
$z(b)\in\2A(|\del_0 b|)\vee\2A(|\del_1 b|)$ for every $b\in\Sigma_1$ such that
$|\del_0 b|\subset\subset|\del_1 b|'$. Thus the arguments in the proof of
\cite[Thm. 3.5]{r4} are applicable despite the fact that we are working in $1+1$
dimensions. We thereby see that there are unique Hilbert spaces 
$H(\2O)\subset\2A(\2O),\ \2O\in\Sigma_0\equiv\2K$ of support $\11$ such that 
$z(b)H(\del_1b)=H(\del_0b)\ \forall b\in\Sigma_1$. Each of these Hilbert spaces
implements an endomorphism $\rho_\2O$ of $\2A$ such that $\rho_\2O\cong\pi$. This
implies that $\rho$ is either reducible or an inner automorphism. \qed\\
\rem This argument needs the split property for double cones. It is not completely 
trivial that the latter follows from the split
property for wedges. It is clear that the latter implies unitary equivalence of 
$\2A(\2O_1)\vee\2A(\2O_2)$ and $\2A(\2O_1)\otimes\2A(\2O_2)$ if
$\2O_1, \2O_2$ are double cones separated by a finite spacelike distance. The split
property for double cones requires more, namely unitary equivalence of 
$\2A(\2O)\vee\2A(\hat{\2O})'$ and $\2A(\2O)\otimes\2A(\hat{\2O})'$ whenever 
$\2O\subset\subset\hat{\2O}$, which is equivalent to the existence of a type I factor $N$
such that $\2A(\2O)\subset N\subset\2A(\hat{\2O})$.
\blemma Let $\2A$ be a local net satisfying Haag duality and the split property for 
wedges. Then the split property for double cones holds. \label{d-split}\elemma
\prf Using the notation of the preceding proof we have
\bea \2A(\2O) &\cong& \2R(W^\2O_R)\otimes\2R(W^\2O_L), \\
   \2A(\hat{\2O}) &\cong& \2R(W^{\hat{\2O}}_R)\otimes\2R(W^{\hat{\2O}}_L) .\eea
By the SPW there are type I factors $N_L, N_R$ such that 
$\2R(W^\2O_L)\subset N_L\subset\2R(W^{\hat{\2O}}_L)$ and 
$\2R(W^\2O_R)\subset N_R\subset\2R(W^{\hat{\2O}}_R)$. Thus 
$Y^{\2O*}(N_R\otimes N_L)Y^\2O$ is a type I factor sitting between $\2A(\2O)$ and
$\2A(\hat{\2O})$. \qed

Having disproved the existence of nontrivial representations localized in double cones
or wedges, we will now prove a result which concerns a considerably larger class of 
representations.
\btheor Let $\2O\mapsto\2A(\2O)$ be a net of observables satisfying Haag duality and
the SPW. Then every irreducible, locally normal representation of the quasilocal algebra
$\2A$ fulfills Haag duality. \label{no-go-2}\etheor
\prf We will show that our assumptions imply those of \cite[Thm. 1]{dri2}.
$\2A$ satisfies the split property for double cones (called `funnel property' in 
\cite{dri2,sum-nps}) by Lemma \ref{d-split}, whereas we also assume condition (1) of 
\cite[Thm.\ 1]{dri2} (Haag duality and irreducibility). Condition (3), which concerns
relative commutants $\2A(\2O_2)\cap\2A(\2O_1)',\ \2O_2\supset\supset\2O_1$ in the vacuum
representation, is an immediate consequence of Prop.\ \ref{relcomm} (we may even 
take $\2O=\2O_1, \2O_2=\2O_3$). Finally, Lemma \ref{a1} implies
\be \2A(\2O)'= \2A(\hat{\2O})'\vee\2A(\2O_L)\vee\2A(\2O_R) ,\ee
where we again use the notation of Fig.\ \ref{f3}. This is more than required by 
Driessler's condition (2). Now \cite[Thm.\ 1]{dri2} applies and we are done. \qed\\
\rems 1. In \cite{sum-nps} a slightly simplified version of \cite[Thm.\ 1]{dri2} is
given which dispenses with condition (2) at the price of a stronger form of condition 
(3). This condition is still (more than) fulfilled by our class of theories.\\
2. Observing that soliton representations are locally normal with respect to both
asymptotic vacua \cite{fre2,schl2}, we conclude at once that Haag duality holds for every
irreducible soliton sector where at least one of the vacua satisfies Haag duality and
the SPW. Consequences of this fact will be explored in the next section. We remark 
without going into details that our results are also of relevance for the construction of
soliton sectors with prescribed asymptotic vacua in \cite{schl2}.

\sectreset{Applications to the Theory of Quantum Solitons}
In \cite{bf} it has been shown that every factorial massive one-particle representation 
(MOPR) in $\ge 2+1$ dimensions is a multiple of an irreducible representation which is 
localizable in every spacelike cone. (Here, MOPR means that the lower bound of the
energy-momentum spectrum consists of a hyperboloid of mass $m>0$ which is separated from
the rest of the spectrum by a mass gap.) In $1+1$ dimensions one is led to irreducible
soliton sectors \cite{fre2} which we will now reconsider in the light of Thms.\ 
\ref{no-go}, \ref{no-go-2}. In this section, where we are concerned with inequivalent
vacuum representations, we will consider a QFT to be defined by a net of abstract 
$C^*$-algebras instead of the algebras in a concrete representation. Given two vacuum
representations $\pi_0^L,\pi_0^R$, a representation $\pi$ is said to be a soliton 
representation of type $(\pi_0^L,\pi_0^R)$ if it is translation covariant and 
\be \pi\restr\2A(W_{L/R})\cong \pi_0^{L/R}\restr\2A(W_{L/R}) ,\label{solit}\ee
where $W_L,\, W_R$ are arbitrary left and right handed wedges, respectively.
An obvious consequence of (\ref{solit}) is local normality of $\pi_0^L,\,\pi_0^R$ with
respect to each other. In order to formulate a useful theory of soliton representations
\cite{fre2} one must assume $\pi_0^{L/R}$ to satisfy wedge duality. After giving a short
review of the formalism in \cite{fre2}, we will show in this section that 
considerably more can be said under the stronger assumption that one of the vacuum 
representations satisfies duality for double cones and the SPW. (Then the other vacuum 
is automatically Haag dual, too.) 

Let $\pi_0$ be a vacuum representation and $W\in\2W$ a wedge. Then by
$\2A(W)_{\pi_0}$ we denote the $W^*$-completion of the $C^*$-algebra $\2A(W)$ with 
respect to the family of seminorms given by
\be \|A\|_T=|\mbox{tr}\,T\pi_0(A)| ,\ee
where $T$ runs through the set of all trace class operators in $\2B(\2H_{\pi_0})$.
Furthermore, we define extensions $\2A_{\pi_0}^L,\,\2A_{\pi_0}^R$ of the quasilocal
algebra $\2A$ by
\be \2A_{\pi_0}^{L/R}=\overline{\bigcup_{W\in\2W_{L/R}} \2A(W)_{\pi_0}}^{\|\cdot\|} ,\ee
where $\2W_L,\2W_R$ are the sets of left and right wedges, respectively. Now, it has 
been demonstrated in \cite{fre2} that, given a $(\pi_0^L,\pi_0^R)$-soliton representation
$\pi$, there are homomorphisms $\rho$ from $\2A_{\pi_0^R}^R$ to $\2A_{\pi_0^L}^R$ such 
that
\be \pi\cong\pi_0^L\circ\rho. \ee
(Strictly speaking, $\pi_0^L$ must be extended to $\2A_{\pi_0^L}^R$, which is trivial 
since $\2A(W)_{\pi_0}$ is isomorphic to $\pi_0(\2A(W))''$.)
The morphism $\rho$ is localized in some right wedge $W$ in the sense that 
\be \rho\restr\2A(W')=\mbox{id}\restr\2A(W') .\ee
Provided that the vacua of two soliton representations $\pi, \pi'$ `fit together'
$\pi_0^R\cong\pi_0^{'L}$ one can define a soliton representation $\pi\times\pi'$ of 
type $\pi_0^L,{\pi_0'}^R$ via composition of the corresponding morphisms:
\be \pi\times\pi'\cong\pi_0^L\circ\rho\rho'\restr\2A .\ee
Alternatively, the entire analysis may be done in terms of left localized morphisms
$\eta$ from $\2A_{\pi_0^L}^L$ to $\2A_{\pi_0^R}^L$. As proved in \cite{fre2}, the 
unitary equivalence class of the composed representation depends neither on the use
of left or right localization nor on the concrete choice of the morphisms.

Whereas for soliton representations there is no analog to the theory of statistics
\cite{dhr3,dhr4,frs1}, there is still a `dimension' $\mbox{ind}(\rho)$ defined by
\be \mbox{ind}(\rho)\equiv[\2A(W)_{\pi_0^L} : \rho(\2A(W)_{\pi_0^R})] ,\ee
where $\rho$ is localized in the right wedge $W$ and $[M:N]$ is the Jones index 
of the inclusion $N\subset M$.
\bprop Let $\pi$ be an irreducible soliton representation such that at least one of the
asymptotic vacua $\pi_0^L, \pi_0^R$ satisfies Haag duality and the SPW. Then $\pi$
and both vacua satisfy the SPW and duality for double cones and wedges. The associated 
soliton-morphism satisfies $\mbox{ind}(\rho)=1$. \eprop
\prf By symmetry it suffices to consider the case where $\pi_0^L$ satisfies HD+SPW.
By Thm.\ \ref{no-go-2} also the representations $\pi$ and $\pi_0^R$ satisfy Haag 
duality since they are locally normal w.r.t.\ to $\pi_0^L$. Let now 
$W_1\subset\subset W_2$ be left wedges. By Prop.\ \ref{wdual}, wedge-duality holds for 
$\pi_0^L$ and $\pi_0^L(\2A(W_1))''\subset\pi_0^L(\2A(W_2))''$ is split. Since
$\pi_0^L(\2A(W_2))''$ is unitarily equivalent to $\pi(\2A(W_2))''$, also 
$\pi(\2A(W_1))''\subset\pi(\2A(W_2))''$ splits.
A fortiori, $\pi$ satisfies the SPW in the sense of Def.\ \ref{spw} and
thus wedge duality by Prop.\ \ref{wdual}. By a similar argument the SPW is carried
over to $\pi_0^R$. Now, for a right wedge $W$ we have:
\be \pi_0^L\circ\rho(\2A(W))^-=\pi_0^L\circ\rho(\2A(W'))'=\pi_0^L(\2A(W'))'=
   \pi_0^L(\2A(W))^- .\ee
By ultraweak continuity on $\2A(W)$ of $\pi_0^L$ and of $\rho$ this implies
\be \rho(\2A(W)_{\pi_0^R})=\2A(W)_{\pi_0^L}, \ee
whence the claim. \qed

This result rules out soliton sectors with infinite index so that \cite[Thm. 3.2]{schl1} 
applies and yields equivalence of the various possibilities of constructing antisoliton 
sectors considered in \cite{schl1}. In particular the antisoliton sector 
is uniquely defined up to unitary equivalence.
Now we can formulate our main result concerning soliton representations.
\btheor Let $\pi_0^L, \pi_0^R$ be vacuum representations, at least one of which satisfies
Haag duality and the SPW. Then all soliton representations of type $(\pi_0^L, \pi_0^R)$
are unitarily equivalent. \etheor
\rem Equivalently, up to unitary equivalence, a soliton representation is completely
characterized by the pair of asymptotic vacua.\\
\prf Let $\pi,\,\pi'$ be irreducible soliton representations of types $(\pi_0,\pi_0')$
and $(\pi_0',\pi_0)$, respectively. They may be composed, giving rise to a soliton
representation of type $(\pi_0,\pi_0)$ (or $(\pi_0',\pi_0')$). This representation
is irreducible since the morphisms $\rho, \rho'$ must be isomorphisms by the proposition.
Now, $\pi\times\pi'$ is unitarily equivalent to $\pi_0$ on left {\it and} right handed 
wedges, which by Thm.\ \ref{no-go} and irreducibility implies 
$\pi\times\pi'\cong\pi_0$. We conclude that every $(\pi_0',\pi_0)$-soliton is an 
antisoliton of every $(\pi_0,\pi_0')$-soliton. This implies the statement of the
theorem since for every soliton representation with finite index there is a corresponding
antisoliton which is unique up to unitary equivalence. \qed\\
\rem The above proof relies on the absence of nontrivial representations which are 
localizable in wedges. Knowing just that DHR sectors do not exist, as follows 
already from Prop.\ \ref{relcomm}, is not enough.

\sectreset{Solitons and DHR Representations of Non-Haag Dual Nets}\label{fixp}
\subsection{Introduction and an Instructive Example}
We have observed that the theory of localized representations of Haag-dual nets of 
observables which satisfy the SPW is trivial. 
There are, however, \qfts\ in $1+1$ dimensions where the net of algebras which is 
most naturally considered as the net of observables does not fulfill Haag duality
in the strong form (\ref{duality}). As mentioned in the introduction,
this is the case if the observables are defined as the fixpoints under 
a global symmetry group of a field net which satisfies (twisted) duality and the SPW. 
The weaker property of wedge duality (\ref{wedge-d}) remains, however. This property is 
also known to hold automatically whenever the local algebras 
arise from a Wightman field theory \cite{biw}. However, for the analysis in 
\cite{dhr3,dhr4,frs2} as well as Sect.\ 4 above one needs full Haag duality. Therefore 
it is of relevance that, starting from a net of observables satisfying only 
(\ref{wedge-d}), one can define a larger but still local net
\be \2A^d(\2O)\equiv\2R(W^\2O_L)\wedge\2R(W^\2O_R) \label{dualnet}\ee
which satisfies Haag duality, whence the name {\it dual net}.
Here $W^\2O_L, W^\2O_R$ are wedges such that $W^\2O_L\cap W^\2O_R=\2O$ and
duality is seen to follow from the fact that the
wedge algebras $\2R(W), W\in\2W$ are the same for the nets $\2A, \2A^d$. (For 
observables arising as group fixpoints the dual net has been computed explicitly in
\cite{mue1}.) It is known \cite{r4,r-aqft} that {\em in $\ge 2+1$ dimensions} 
representations $\pi$ satisfying the DHR criterion (\ref{dhr}) extend uniquely to DHR 
representations $\hat{\pi}$ of the (appropriately defined) dual net. Furthermore, the 
categories of DHR representations of $\2A$ and $\2A^d$, respectively, and their 
intertwiners are isomorphic. Thus, instead of $\2A$ one may as well study $\2A^d$ to 
which the usual methods are applicable. (The original net is needed only to satisfy 
essential duality, which is implied by wedge duality.)
In $1+1$ dimensions things are more complicated.
As shown in \cite{r1} there are in general two different extensions 
$\hat{\pi}^L,\hat{\pi}^R$. They coincide iff one (thus both) of them is a DHR 
representation. Even before defining precisely these extensions we can state
the following consequence of Thm.\ \ref{no-go}.
\bprop Let $\2A$ be a net of observables satisfying wedge duality and the SPW. Let
$\pi$ be an irreducible DHR or wedge representation of $\2A$ which is not unitarily 
equivalent to the defining (vacuum) representation. Then there is no extension 
$\hat{\pi}$ to the dual net $\2A^d$ which is still localized in the DHR or wedge sense. 
\eprop
\prf Assume $\pi$ to be the restriction to $\2A$ of a wedge-localized representation 
$\hat{\pi}$ of $\2A^d$. As the latter is known to be either reducible or unitarily 
equivalent to $\pi_0$, the same holds for $\pi$. This is a contradiction. \qed\\
The fact that the extension of a localized representation of $\2A$ to the dual net
$\2A^d$ cannot be localized, too, partially undermines the original motivation for 
considering these extensions. Nevertheless, one may entertain the hope that there is
something to be learnt which is useful for a model-independent analysis of the
phenomena observed in models. Before we turn to the general examination
of the extensions $\hat{\pi}^L,\hat{\pi}^R$ we consider the most instructive example.

It is provided by the fixpoint net under an unbroken global symmetry group of a field net
as studied in \cite{mue1}. We briefly recall the framework. Let $\2O\mapsto\2F(\2O)$ be 
(for simplicity) bosonic, i.e.\ local, net of von Neumann algebras acting on the Hilbert 
space $\2H$ and satisfying Haag duality and the SPW. On $\2H$ there 
are commuting strongly continuous representations of the \poinc\ group and of a group 
$G$ of inner symmetries. Both groups leave the vacuum $\Omega$ invariant.
Defining the fixpoint net 
\be \2A(\2O)=\2F(\2O)^G=\2F(\2O)\cap U(G)' \ee
 and its restriction
\be \6A(\2O)=\2A(\2O)\restr\2H_0 \ee
to the vacuum sector (= subspace of $G$-invariant vectors)
we consider $\6A(\2O)$ as the observables. It is well known that the net $\6A$ satisfies
only wedge duality. Nevertheless, one very important result of \cite{dhr1} remains true,
namely that the restrictions of $\2A$ to the charged sectors $\2H_\chi$ which are labeled
by the characters $\chi\in\hat{G}$, interpreted as representations of the abstract 
$C^*$-algebra $\2A$, satisfy the DHR criterion and are connected to the vacuum by 
charged fields. I.e.\ the representation of $\2A$ in $\2H_\chi$ is of the form
\be \pi_\chi(A)=A\restr\2H_\chi\cong\pi_\chi^\2O(A)=\psi\,A\,\psi^*\restr\2H_0 ,
\label{pichi}\ee
where $\psi\in\2F(\2O)$ and $\alpha_g(\psi)=\chi(g)\psi$.

It was shown in \cite[Thm. 3.10]{mue1} that the dual net in the vacuum sector is given by
\be \6A^d(\2O)=\hat{\2A}_L(\2O)\restr\2H_0=\hat{\2A}_R(\2O)\restr\2H_0, \ee
where
\be \hat{\2A}_{L/R}(\2O)=\hat{\2F}_{L/R}(\2O)^G=\hat{\2F}_{L/R}(\2O)\cap U(G)'. \ee
Here the nonlocal nets $\hat{\2F}_{L/R}(\2O)$ are obtained by adjoining to $\2F(\2O)$
the {\it disorder operators} \cite{mue1} $U^\2O_L(G)$ or $U^\2O_R(G)$, respectively,
which satisfy
\be\ba{ccccc} \mbox{Ad}\,U_L^\2O(g)\restr\2F(W^\2O_{LL}) &=& \alpha_g &=&
   \mbox{Ad}\,U_R^\2O(g)\restr\2F(W^\2O_{RR}) ,\\
   \mbox{Ad}\,U_L^\2O(g)\restr\2F(W^\2O_{RR}) &=& \mbox{id} &=&
   \mbox{Ad}\,U_R^\2O(g)\restr\2F(W^\2O_{LL})  
\ea\label{disord}\ee
and transform covariantly under the global symmetry:
\be U(g)\,U^\2O_{L/R}(h)\,U(g)^*=U^\2O_{L/R}(ghg^{-1}). \ee

For the moment we restrict to the case of abelian groups $G$. The disorder operators 
commuting with $G$, $\hat{\2A}_{L/R}(\2O)$ is simply $\2A(\2O)\vee U^\2O_{L/R}(G)''$. On
the $C^*$-algebras $\hat{\2A}_L$ and $\hat{\2A}_R$ there is an action of the dual group
$\hat{G}$ which acts trivially on $\2A$ and via 
\be \hat{\alpha}_\chi(U^\2O_{L/R}(g))=\chi(g)\,U^\2O_{L/R}(g) \quad\forall\2O\in\2K\ee 
on the disorder operators. Since this action commutes with the \poinc\ group
and since it is spontaneously broken 
($\omega_0\circ\hat{\alpha}_\chi\ne\omega_0\ \forall\chi\ne e_{\hat{G}}$) it 
gives rise to inequivalent vacuum states on $\hat{\2A}$ via 
\be \omega_\chi=\omega_0\circ\hat{\alpha}_\chi.\ee

The extensions $\hat{\pi}_{\chi,L}, \hat{\pi}_{\chi,R}$ of $\pi_\chi$ to the dual net 
$\2A^d$ can now defined using the right hand side of (\ref{pichi}) by allowing $A$ to
be in $\hat{\2A}_L$ or $\hat{\2A}_R$. As is obvious from the commutation relation 
(\ref{disord}) between fields and disorder operators, the extension $\hat{\pi}_{\chi,L}$ 
($\hat{\pi}_{\chi,R}$) is nothing but a soliton sector interpolating between the vacua 
$\omega_0$ and $\omega_{\chi^{-1}}$ ($\omega_\chi$ and $\omega_0$). The moral is that
the net $\2A^d$, while not having nontrivial localized representations by Thm.\ 
\ref{no-go}, admits soliton representations. Furthermore, with respect to $\2A^d$,
the charged fields $\psi_\chi$ are creation operators for solitons since they
intertwine the representations of $\2A^d$ on $\2H_0$ and $\2H_\chi$.
Due to $U^\2O_L(g)\,U^\2O_R(g)=U(g)$ and $U(g)\restr\2H_\chi=\chi(g)\11$ we have
\be U^\2O_L(g)\restr\2H_\chi=\chi(g)\,U^\2O_R(g^{-1})\restr\2H_\chi ,\ee
so that the algebras $\hat{\2A}_{L/R}(\2O)\restr\2H_\chi$ are independent of whether we
use the left or right localized disorder operators. 
In particular, in the vacuum sector $U^\2O_L(g)$ and $U^\2O_R(g^{-1})$ coincide, but
due to the different localization properties it is relevant whether $U^\2O_L(g)$,
considered as an element of $\2A^d$, is represented on $\2H_\chi$ by $U^\2O_L(g)$ or by 
$\chi(g)\,U^\2O_R(g^{-1})$. This reasoning shows that the two 
possibilities for extending a localized representation of a general non-dual net to 
a representation of the dual net correspond in the fixpoint situation at hand to the 
choice between the nets $\hat{\2A}_L$ and $\hat{\2A}_R$
arising from the field extensions $\hat{\2F}_L$ and $\hat{\2F}_R$.

\subsection{General Analysis}
We begin by first assuming only that $\pi$ is localizable in wedges.
Let $\2O$ be a double cone and let $W_L, W_R$ be left and right handed wedges,
respectively, containing $\2O$. By assumption the restriction of $\pi$ to 
$\2A(W_L), \2A(W_R)$ is unitarily equivalent to $\pi_0$. Choose unitary implementers
$U_L, U_R$ such that
\be\ba{ccc} Ad\,U_L\restr\2A(W_L) &=& \pi\restr\2A(W_L), \\
   Ad\,U_R\restr\2A(W_R) &=& \pi\restr\2A(W_R). \ea\ee
Then $\hat{\pi}^L, \hat{\pi}^R$ are defined for $A\in\2A^d(\2O)$ by
\be\ba{ccc} \hat{\pi}^L (A) &=& U_L \, A\, U_L^* , \\
   \hat{\pi}^R (A) &=& U_R \, A\, U_R^*. \ea\label{pihat}\ee 
Independence of these definitions of the choice of $W_L, W_R$ and the implementers
$U_L, U_R$ follows straightforwardly from wedge duality. We state some immediate
consequences of this definition.
\bprop $\hat{\pi}^L,\hat{\pi}^R$ are irreducible, locally normal representations of 
$\2A^d$ and satisfy Haag duality. $\hat{\pi}^L,\hat{\pi}^R$ are normal on left and right
handed wedges, respectively. \eprop
\prf Irreducibility is a trivial consequence of the assumed irreducibility of $\pi$
whereas local normality is obvious from the definition (\ref{pihat}). Thus, Thm.\ 
\ref{no-go-2} applies and yields Haag duality in both representations. Normality of, say,
$\hat{\pi}^L$ on left handed wedges $W$ follows from the fact that we may use the 
same auxiliary wedge $W_L\supset W$ and implementer $U_L$ for all double cones 
$\2O\subset W$. \qed

Clearly, the extensions $\hat{\pi}^L,\hat{\pi}^R$ cannot be normal w.r.t.\ $\pi_0$ on
right and left wedges, respectively, for otherwise Thm.\ \ref{no-go} would imply
unitary equivalence to $\pi_0$. In general, we can only conclude localizability
in the following weak sense. Given an arbitrary left handed wedge $W, \ \hat{\pi}^L$
is equivalent to a representation $\rho$ on $\2H_0$ such that 
$\rho(A)=A\ \forall A\in\2A(W)$. Furthermore, by duality $\rho$ is an isomorphism of 
$\2A(W')$ onto a weakly dense subalgebra of $\2R(W')$ which is only continuous in the 
norm. In favorable cases like the one considered above
this is a local symmetry, acting as an automorphism of 
$\2A(W')$. But we will see shortly that there are perfectly non-pathological situations
where the extensions are not of this particularly nice type. In complete generality, the
best one can hope for is normality with respect to another
vacuum representation $\pi_0'$. In particular, this is automatically the case if $\pi$ 
is a massive one-particle representation \cite{bf} which we did not assume so far.

If the representation $\pi$ satisfies the DHR criterion, i.e.\ is localizable in
double cones, we can obtain stronger results concerning the localization properties
of the extended representations $\hat{\pi}_L,\hat{\pi}_R$. By the criterion,
there are unitary operators $X^\2O: \2H_\pi\to\2H_0$ such that
\be \pi^\2O(A)\equiv X^\2O\,\pi(A)\,X^{\2O*}=A\quad \forall A\in\2A(\2O') .\ee
(By wedge duality, $X^\2O$ is unique up to left multiplication by operators in
$\2A^d(\2O)$.) Considering the representations 
\be \hat{\pi}^\2O_{L/R} = X^\2O\hat{\pi}_{L/R}X^{\2O*} \ee
on the vacuum Hilbert space $\2H_0$, it is easy to verify that 
\bea \hat{\pi}^\2O_L\restr\2A^d(W^\2O_{LL}) &=& \mbox{id}\restr\2A^d(W^\2O_{LL}), 
  \label{pihatl}\\
   \hat{\pi}^\2O_R\restr\2A^d(W^\2O_{RR}) &=& \mbox{id}\restr\2A^d(W^\2O_{RR}). \eea
We restrict our attention to $\hat{\pi}^\2O_L$, the other extension behaving similarly.
If $A\in\2A(\tilde{\2O})$ then $\hat{\pi}_L(A)={X^{\2O_r}}^*\,A\,X^{\2O_r}$ whenever 
$\2O_r>\tilde{\2O}$. Therefore
\be \hat{\pi}^\2O_L(A)=X^\2O{X^{\2O_r}}^*\,A\,X^{\2O_r}X^{\2O*} ,\label{pihat2}\ee
where the unitary $X^\2O{X^{\2O_r}}^*$ intertwines $\pi^\2O$ and $\pi^{\2O_r}$. 
Associating with every pair $(\2O_1,\2O_2)$ two other double cones by
\bea \hat{\2O} &=& \sup(\2O_1,\2O_2), \\
   \2O_0 &=& \hat{\2O}\cap\2O_1'\cap\2O_2' \eea
($\2O_0$ may be empty) and defining
\be \2C(\2O_1,\2O_2)=\2A^d(\hat{\2O})\cap\2A(\2O_0)' ,\ee
we can conclude by wedge duality that
\be X^\2O{X^{\2O_r}}^*\in\2C(\2O,\2O_r) .\label{m-lok}\ee
Thus $\hat{\pi}^\2O_L(A)$ as given by (\ref{pihat2}) is contained in 
$\2A^d(\sup(\2O,\tilde{\2O},\2O_r))$ which already shows that $\hat{\pi}^\2O_L$ maps
the quasilocal algebra $\2A^d$ into itself (this does not follow if $\pi$ is only 
localizable in wedges). Since the double cone $\2O_r>\2O$ may be chosen arbitrarily 
small and appealing to outer regularity of the dual net $\2A^d$ we even have
$\hat{\pi}^\2O_L(A)\in\2A^d(\sup(\2O,\tilde{\2O}))$ and thus finally
\be \hat{\pi}^\2O_L(\2A^d(\tilde{\2O}))\subset\2C(\2O,\tilde{\2O}) .\label{deloc}\ee
This result has two important consequences. Firstly, it implies that the representation
$\hat{\pi}^\2O_L$ maps the quasilocal algebra into itself:
\be \hat{\pi}^\2O_L(\2A^d)\subset\2A^d .\ee
This fact is of relevance since it allows the extensions 
$\hat{\pi}^\2O_{1,L}, \hat{\pi}^\2O_{2,L}$ of two DHR representations $\pi_1, \pi_2$
to be composed in much the same way as the endomorphisms of $\2A$ derived from DHR
representations in
the Haag dual case. In this respect, the extensions $\hat{\pi}_{L/R}$ are better
behaved than completely general soliton representations as studied in \cite{fre2}.

The second consequence of (\ref{deloc}) is that the representations
$\hat{\pi}^\2O_L$ (and $\hat{\pi}^\2O_R$), while still mapping local algebras into 
local algebras, may deteriorate the localization.
We will see below that this phenomenon is not just a theoretical possibility but
really occurs. Whereas one might hope that one could build a DHR theory for non-dual nets
upon the endomorphism property of the extended representations, their weak localization
properties and the inequivalence of $\hat{\pi}_L$ and $\hat{\pi}_R$ seem to constitute
serious obstacles. It should be emphasized that the above considerations owe a lot to 
Roberts' local 1-cohomology \cite{r1,r4,r-aqft}, but (\ref{deloc}) seems to be new.

\subsection{Fixpoint Nets: Non-abelian Case}
We now generalize our analysis of fixpoint nets to nonabelian (finite) groups $G$, where 
the outcome is less obvious a priori.
Let $\hat{A}=\sum_{g\in G}F_g U^{\tilde{\2O}}_L(g)\in\hat{\2A}_L(\tilde{\2O_1})$ 
($F_g$ must satisfy the condition given in \cite[Thm. 3.16]{mue1}) and let 
$\psi_i\in\2F(\2O_2)$,
where $\2O_2<\2O_1$ (i.e.\ $\2O_2\subset W^{\2O_1}_{LL}$) be a multiplet of 
field operators transforming according to a 
finite dimensional representation of $G$. Then
\be \sum_i \psi_i \ (\sum_{g\in G} F_gU^{\2O_1}_L(g)) \ \psi_i^* = 
   \sum_{g\in G} (\sum_i \psi_i\alpha_g(\psi_i^*)) \, F_gU^{\2O_1}_L(g) .
\label{psihat}\ee
In contrast to the abelian case where $\psi\alpha_g(\psi^*)$ is just a phase, 
$O_g\equiv\sum_i \psi_i\alpha_g(\psi_i^*)$ is a nontrivial unitary operator 
\be O_g^{-1} = O_g^*=\sum_i \alpha_g(\psi_i)\psi_i^* \ee
satisfying
\be \alpha_k(O_g)=O_{kgk^{-1}} .\ee
In particular (\ref{psihat}) is not contained in $\2A^d(\2O_1)$ which implies that 
the map $\hat{A}\mapsto\sum_i\psi_i\,\hat{A}\,\psi_i^*$ does not reduce to a 
local symmetry on $\hat{\2A}_L(W^{\2O_2}_{RR})$. Rather, we obtain a monomorphism
into $\hat{\2A}_L(W^{\2O_2}_R)$. Defining $\hat{\2O}$ and $\2O_0$ as above we clearly see
that (\ref{psihat}) is contained in $\2A^d(\hat{\2O})$. Furthermore, due to the relative 
locality of the net $\2A$ with respect to $\2A^d$ and $\2F$, (\ref{psihat}) commutes
with $\2A(\2O_0)$. Thus we obtain precisely the localization properties which were
predicted by our general analysis above.

We close this section with a discussion of the duality properties in the 
extended representations $\hat{\pi}$. In the case of abelian groups $G$ Haag duality
holds in all charged sectors since these are all simple. Our abstract result 
in Thm.\ \ref{no-go-2} to the effect that duality obtains in {\it all} locally
normal irreducible representations of the dual net applies, of course, to the situation 
at hand. We conclude that Haag duality also holds for the non-simple sectors which by
necessity occur for non-abelian groups $G$. Since this result is somewhat 
counterintuitive (which explains why it was overlooked in \cite{mue1}) we verify it
by the following direct calculation.
\blemma The commutants of the algebras $\hat{\2A}_L(\2O)$ are given by
\be \hat{\2A}_L(\2O)'=\hat{\2A}_L(W^\2O_{LL})\vee\hat{\2F}_L(W^\2O_{RR}) \quad
  \forall\2O\in\2K .\label{comm}\ee
\elemma
\prf For simplicity we assume $\2F$ to be a local net for a moment. Then
\bea \hat{\2A}_L(\2O)' &=& (\hat{\2F}_L(\2O)\wedge U(G)')' = \hat{\2F}_L(\2O)'\vee U(G)'' \nn\\
  &=& (\2F_L(\2O)\vee U^\2O_L(G)'')'\vee U(G)''=(\2F_L(\2O)'\wedge U^\2O_L(G)')\vee U(G)'' 
\nn\\
  &=& ((\2F_L(W^\2O_{LL})\vee\2F_L(W^\2O_{RR}))\wedge U^\2O_L(G)')\vee U(G)'' \\
  &=& (\2F_L(W^\2O_{LL})\wedge U^\2O_L(G)')\vee\2F_L(W^\2O_{RR})\vee U(G)'' \nn\\
  &=& \hat{\2A}_L(W^\2O_{LL})\vee\hat{\2F}_L(W^\2O_{RR}) .\nn\eea
The fourth line follows from the third using the split property.
In the last step we have used the identities $\hat{\2A}_L(W_L)=\2A_L(W_L)$ and
$\2F_L(W_R)\vee U(G)''=\hat{\2F}_L(W_R)$ which hold for all left (right) handed wedges 
$W_L$ ($W_R$), cf. \cite[Prop. 3.5]{mue1}. Now, if $\2F$ satisfies twisted duality,
(2.23) of \cite{mue1} leads to 
$\2F(\2O)\vee U_L^\2O(G)''\cong\2F(W^\2O_R)\vee U(G)''\otimes\2F(W^\2O_L)$ and
$(\2F(\2O)\vee U_L^\2O(G)'')'\cong\2A(W^\2O_R)\otimes\2F(W^\2O_{RR})^t$. Using this it
is easy to verify that (\ref{comm}) is still true. \qed
\bprop The net $\hat{\2A}_L$ satisfies Haag duality in restriction to every invariant 
subspace of $\2H$ on which $\hat{\2A}_L$ acts irreducibly. \eprop
\prf We recall that the representation $\pi$ of $\hat{\2A}_{L/R}$ on $\2H$ is of the 
form $\pi=\oplus_{\xi\in\hat{G}} \,d_\xi\pi_\xi$. Let thus $P$ be an orthogonal 
projection onto a subspace $\2H_\xi\subset\2H$ on which $\hat{\2A}_L$ acts as the
irreducible representation $\pi_\xi$. Since $P$ commutes with 
$\2A_L(\2O)$ and $\2A_L(W^\2O_{LL})$ we have
\bea P\,\hat{\2A}_L(\2O)'\,P &=&
   P\,\hat{\2A}_L(W^\2O_{LL})\vee\hat{\2F}_L(W^\2O_{RR})\,P \nn\\
  &=& \hat{\2A}_L(W^\2O_{LL})\vee (P\,\hat{\2F}_L(W^\2O_{RR})\,P) \\
  &=& P\, \hat{\2A}_L(W^\2O_{LL})\vee \hat{\2A}_L(W^\2O_{RR})\,P ,\nn\eea
which implies
\be (\hat{\2A}_L(\2O)\restr\2H_\xi)' = \hat{\2A}_L(W^\2O_{LL})\vee 
  \hat{\2A}_L(W^\2O_{RR})\restr\2H_\xi .\ee
\noindent\qed\\
This provides a concrete verification of Thm.\ \ref{no-go-2} in a special, albeit
important situation.

\sectreset{Conclusions and Outlook}
We have seen that the combination of Haag duality with the split property for wedges 
has remarkable unifying power. It implies factoriality of the double cone algebras, 
$n$-regularity for all $n$ and irreducibility of time-slice algebras. As a consequence
of the minimality of relative commutants of double cone algebras we obtain Haag duality 
in all irreducible, locally normal representations. The strongest result concerns
the absence not only of locally generated superselection (DHR) sectors but also of 
charges localized in wedges. This in turn implies the uniqueness up to unitary 
equivalence of soliton sectors with prescribed asymptotic vacua. In the following we 
briefly relate these results to what is known in concrete models in $1+1$ dimensions.\\
a) {\it The free massive scalar field.}\ \ Since this model is known to 
satisfy Haag duality and the SPW, Thm.\ \ref{no-go} constitutes a high-brow proof
of the well known absence of local charges. Furthermore, there are no 
non-trivial soliton sectors, since the vacuum representation is unique \cite{segal}. 
Thus, the irreducible representations constructed in \cite{str}, which 
are inequivalent to the vacuum, must be rather pathological. In fact, they are
equivalent to the (unique) vacuum only on left wedges.\\
b) $\2P(\phi)_2$-{\it models.} These models have been shown \cite{dri2} to satisfy Haag 
duality in all pure phases, but there is no proof of the SPW. Yet, the split property for
double cones, the minimality of relative commutants and strong additivity, thus also
the time slice property, follow immediately from the corresponding properties for
the free field via the local Fock property. These facts already imply the non-existence
of DHR sectors and Haag duality in all irreducible locally normal sectors.
All these consequences are compatible with the conjecture that the SPW holds. 
There seems, however, not to be a proof of the absence of wedge sectors.\\
c) {\it The sine-Gordon/Thirring model.} For this model neither Haag duality
nor the SPW are known. In the case $\beta^2=4\pi$, however, for which the SG
model corresponds to the free massive Dirac field, there seems to be no doubt 
that the net $\hat{\2A}$ constructed like in Sect.\ \ref{fixp} from the free Dirac field
is exactly the local net of the SG model. As shown in \cite{mue1}, also 
$\hat{\2A}$ satisfies Haag duality and the SPW. Since from the point of view
of constructive QFT there is nothing special about $\beta^2=4\pi$ one may 
hope that both properties hold for all $\beta\in[0,8\pi)$. 

In view of the results of this paper as well as of \cite{mue1} it is highly desirable to 
clarify the status of the SPW in interacting massive models like b) and c) as well as 
that of Haag duality in case c). (Also the Gross-Neveu model might be expected to satisfy
both assumptions.) The most promising approach to this problem should be identifying
conditions on a set of Wightman (or Schwinger) distributions which imply Haag duality 
and the SPW, respectively, for the net of algebras generated by the fields. For a first 
step in this direction see \cite[Sect.\ IIIB]{buchh2}. \\ \\ 
\noindent{\it Acknowledgments.} I am greatly indebted to K.-H.~Rehren for his interest 
and encouragement, many helpful discussions and several critical readings of the
manuscript. Conversations with K.~Fredenhagen, J.~Roberts, B.~Schroer, and 
H.-W.~Wiesbrock are gratefully acknowledged. The work was completed at the Erwin 
Schr\"odinger Institute, Vienna which kindly provided hospitality and financial support.
Last but not least, I thank P.~Croome for a very thorough proofreading of a
preliminary version.

\end{document}